\documentclass{kluwer}    

\newdisplay{guess}{Conjecture}
\usepackage{graphicx}
\begin{document}
\begin{article}
\begin{opening}
\title{Observational tests of the galaxy formation process}
\author{Gianfranco \surname{De Zotti} and Gian Luigi
\surname{Granato}} \institute{INAF - Osservatorio Astronomico di
Padova \& SISSA, Trieste, Italy}
\author{Laura \surname{Silva}} \institute{INAF - Osservatorio Astronomico di
Trieste, Italy} \author{Luigi \surname{Danese}} \institute{SISSA,
Trieste, Italy}\runningauthor{G. De Zotti et al.}
\runningtitle{Observational tests of the galaxy formation process}
\date{December 2, 2003}

\begin{abstract}
The mutual feedback between star formation and nuclear activity in
large spheroidal galaxies may be a key ingredient to overcome
several difficulties plaguing current semi-analytic models for
galaxy formation. We discuss some observational implications of
the model by Granato et al. (2003) for the co-evolution of
galaxies and active nuclei at their centers and stress the
potential of the forthcoming surveys of the Sunyaev-Zeldovich
effect on arcminute scales, down to $\mu$K levels, to investigate
the early galaxy formation phases, difficult to access by other
means.
\end{abstract}

\end{opening}

\section{Introduction}

As first pointed out by Silva (1999), and confirmed by other
groups (e.g., Devriendt \& Guiderdoni, 2000; Kaviani et al.,
2003), semi-analytic models, built in the framework of the
currently standard hierarchical clustering paradigm, under-predict
by a large factor the surface density of galaxies detected by
(sub)-mm surveys. The available redshift information (Dunlop,
2001; Chapman et al., 2003) albeit limited, indicates a strong
excess of high luminosity, presumably very massive, galaxies at
high $z$, compared to model predictions. The excess is confirmed
by deep IR (I- and K-band) surveys (Daddi et al., 2003; Kashikawa
et al., 2003; Poli et al., 2003; Pozzetti et al., 2003; Somerville
et al., 2003).

Yet another difficulty for these models comes the observed
colour-magnitude and [$\alpha/Fe$]-magnitude relations for
elliptical galaxies; in particular, the $\alpha/Fe$ ratio
increases with luminosity (Thomas et al., 2002). Since $Fe$ is
mostly produced by type Ia supernovae, the over-abundance of
$\alpha$-elements, compared to $Fe$, is most simply understood if
the star formation is halted before type Ia SNe can enrich the
interstellar medium. But this interpretation implies a shorter
duration of the star-formation activity for more luminous -- i.e.
more massive -- galaxies (Granato et al. 2001; Romano et al.,
2002, and references therein). On the other hand, the depth of the
potential well increases with the galaxy mass. Therefore,
supernova explosions can unbind the interstellar medium (thus
halting the star formation) more easily in less massive galaxies.
This is why current semi-analytic models (Devriendt \& Guiderdoni,
2000; Cole et al., 2000; Somerville et al., 2001; Menci et al.,
2002) tend to predict an [$\alpha/Fe$]-magnitude relation quite
opposite to the observed one.

However, the data summarized above are not necessarily in
contradiction with the hierarchical clustering scenario, since the
predicted number of large dark matter halos at the relevant
redshifts is consistent with that of observed luminous galaxies.
But the observed high-$z$ galaxies are far more luminous, and
their chemical abundances are very different, than expected if
most of the star formation occurs in small objects that later
merged to build the large spheroidal galaxies. It is thus likely
that the inconsistencies originate from a poor modelling of the
physical processes involved.

The observations apparently require that large halos present at
high redshift have essentially all their baryons still in a
gaseous form and give origin to a single gigantic burst of star
formation, which is halted by an energy injection that, unlike
that from supernovae, is increasingly effective with increasing
halo mass. As discussed by Granato et al. (2001, 2003), a very
important ingredient that has been largely ignored in
semi-analytic models is the mutual feedback between star-formation
and nuclear activity of spheroidal galaxies. In the following, we
will summarize the main aspects of the model by Granato et al.
(2003), with emphasis on the possible role of the feedback, and
present some observational tests of the ensuing galaxy formation
scenario.

\section{Co-evolution of galaxies and active nuclei}

According to the model by Granato et al. (2003) the evolution of
the gas within massive dark matter halos, forming at the rate
predicted by the canonical hierarchical clustering scenario, is
controlled by gravity, radiative cooling, and heating by feedback
from supernovae and from the growing active nucleus. Supernova
heating is increasingly effective with decreasing binding energy
in slowing down the star formation and in driving gas outflows.
Thus, the more massive proto-galaxies virializing at earlier times
are the sites of the faster star-formation. The correspondingly
higher radiation drag fastens the angular momentum loss by the
gas, resulting in a larger accretion rate onto the central
black-hole.

This scenario has a clear-cut implication for the relationship
among the black-hole mass and the stellar velocity dispersion,
which has been interpreted by many authors in terms of either
energy or momentum balance of the quasar-driven outflow (Silk \&
Rees, 1998; Fabian, 1999; Cavaliere et al., 2002; King, 2003).
These approaches yield a power-law relationship, $M_{\rm BH}
\propto \sigma^p$, with $p$ in the range 4 to 5. On the other
hand, according to Granato et al. (2003) for lower and lower halo
masses an increasing fraction of mechanical energy deposited in
the interstellar medium does not come from nuclear outflows but
from supernovae. The heating due to the latter is increasingly
effective in slowing down the black-hole growth in the less
massive halos. Therefore, while the power-law $M_{\rm
BH}$--$\sigma$ relationship is recovered (with both slope and
normalization nicely consistent with observations) for $\log\sigma
\gsim 2.2$, a steepening is predicted for lower values (see Fig.~6
of Granato et al., 2003).


The observed $M_{\rm BH}$--$\sigma$ relation entails an increase
of the $M_{\rm BH}/M_{\rm halo}$ ratio with increasing $M_{\rm
halo}$. Therefore the gas heating per unit halo mass increases
with halo mass, if the quasar radiates at the Eddington limit (as
it is likely during its fast growth phase) and releases a constant
fraction of its power in mechanical form. However, this is not
enough to sweep out the interstellar gas of the most massive
galaxies on a time-scale $\leq 5\times 10^8$--$10^9$ years, i.e.
as short as necessary to avoid the $Fe$ enrichment by type Ia
supernovae. To this end, it is necessary that the fraction of
mechanical energy released by the quasar is either higher or more
tightly coupled to the interstellar gas for more massive galaxies.
The latter is a likely possibility: if anything, the optical depth
is larger for bigger galaxies. However, the physics of such
coupling is poorly understood. Granato et al. (2003), based on the
model for AGN-driven outflows by Murray et al. (1995), assume that
the fraction, $f_h$, of bolometric luminosity going into gas
heating increases as $L_{\rm bol}^{1/2}$. Interestingly, the
effective, luminosity weighted, value of $f_h$ is consistent with
the value ($\sim 0.1$) that can account for the pre-heating of the
intergalactic medium in groups of galaxies (Cavaliere et al.,
2002; Platania et al., 2002).

This simple recipe yields a duration, $T_{sb}$, of the most active
star formation phase decreasing with increasing $M_{\rm halo}$ and
virialization redshift $z_{\rm vir}$. In the redshift range
$3\lsim z_{\rm vir} \lsim 6$, $T_{\rm sb} \simeq 0.5$--$1\,$Gyr
for $M_{\rm halo} \geq {\rm few}\times 10^{12}\, M_{\odot}$. 
Thus, the physical processes acting on baryons reverse the
order of the formation of spheroidal galaxies with respect to the
hierarchical assembly of dark matter halos ({\it Anti-hierarchical
Baryon Collapse} scenario; Granato et al., 2001).

Coupling the model with GRASIL (Silva et al., 1998), the code
computing in a self-consistent way the chemical and
spectrophotometric evolution of galaxies over a very wide
wavelength interval, Granato et al. (2003) obtained predictions in
excellent agreement with observations for a number of observables
which proved to be extremely challenging for all the current
semi-analytic models, including the sub-mm counts and the
corresponding redshift distributions, and the epoch-dependent
K-band luminosity function of spheroidal galaxies.

\section{Sunyaev-Zeldovich effects from the early phases of galaxy formation}

The model by Granato et al. (2003) effectively assumes that the
gravitational potential wells appear instantaneously.
This stems from the analysis by Zhao et al. (2003) of
high-resolution N-body simulations, showing that the growth of
dark matter halos consists of an early fast accretion phase, which
establishes the potential well, followed by a slow accretion
phase, when the mass increases by a substantial factor without
changing the potential well significantly. On the other hand,
according to other analyses (e.g., Lacey \& Cole, 1993) the halo
(and the associated potential well) is assembled continuously over
a long time-scale.

Furthermore, the model makes the usual assumption that the gas is
shock-heated to the virial temperature as soon as the potential
well develops and is held in quasi-static equilibrium while it
cools and contracts (Rees \& Ostriker, 1977; White \& Rees, 1978).
This is also being disputed: according to some recent analyses
(Birnboim \& Dekel, 2003; Binney, 2003) only a fraction of the gas
is heated.

De Zotti et al. (2003) pointed out that galactic scale
Sunyaev-Zeldovich (SZ) effects may be a viable tool to investigate
these early phases (see also Rosa-Gonzalez et al., 2003). In fact,
the forthcoming generation of SZ instruments will be orders of
magnitude more efficient, allowing to reach $\mu$K level with
arcminute resolution (Carlstrom et al., 2002). The proto-galactic
gas in a large galaxy may be expected to have a large thermal
energy content, leading to a SZ signal at the several $\mu$K level
during two evolutionary phases: i) when the protogalaxy collapses
{\bf if} the gas is shock-heated to the virial temperature; ii)in
a later phase, as the result of strong feedback from a flaring
active nucleus (Natarajan \& Sigurdssson, 1999; Aghanim et al.,
2000; Platania et al., 2002; Lapi et al., 2003).

\begin{figure}[t]
\tabcapfont
\centerline{%
\begin{tabular}{c@{\hspace{0pc}}c}
\includegraphics[width=2.5in, height=3in]{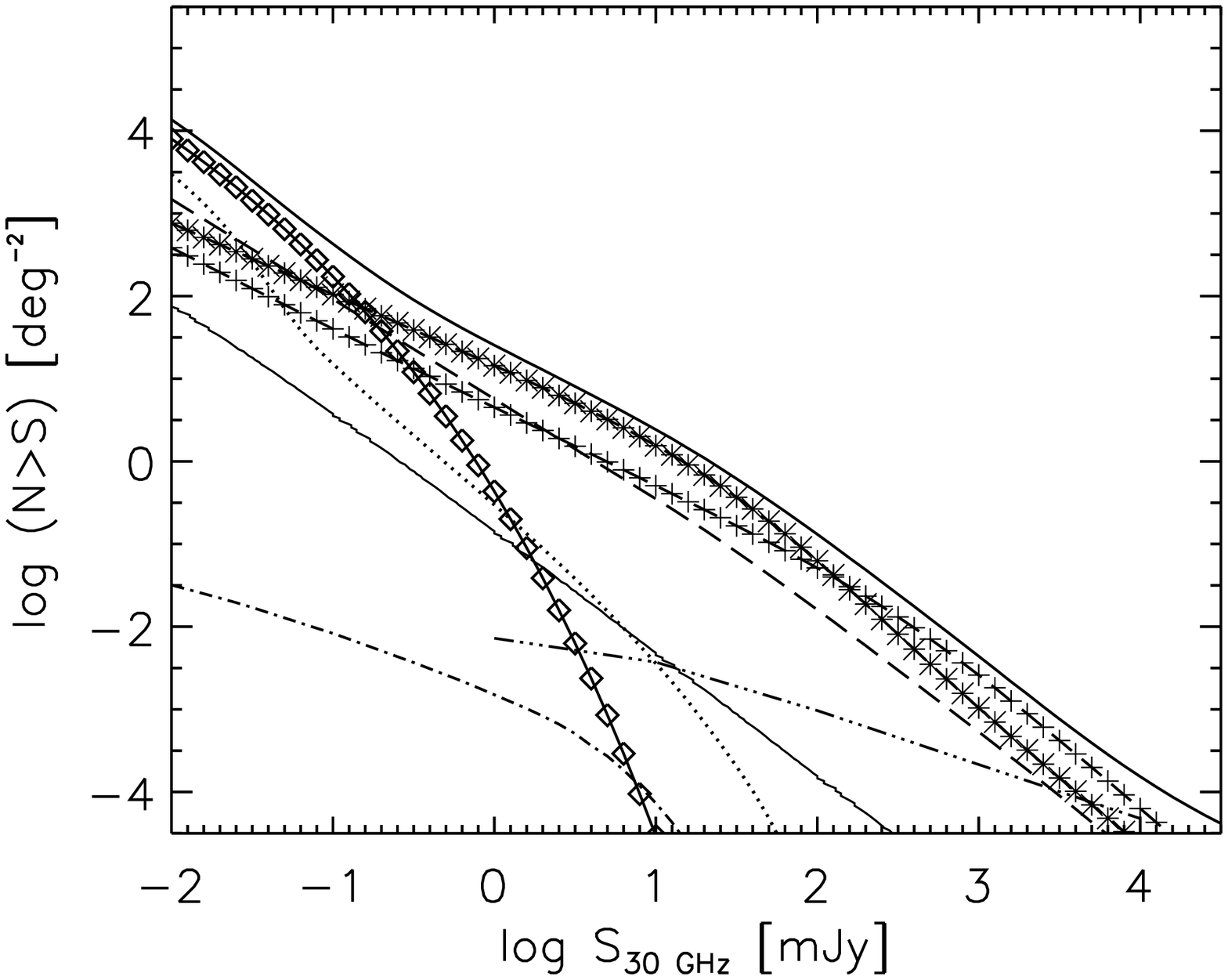} &
\includegraphics[width=2.5in, height=3in]{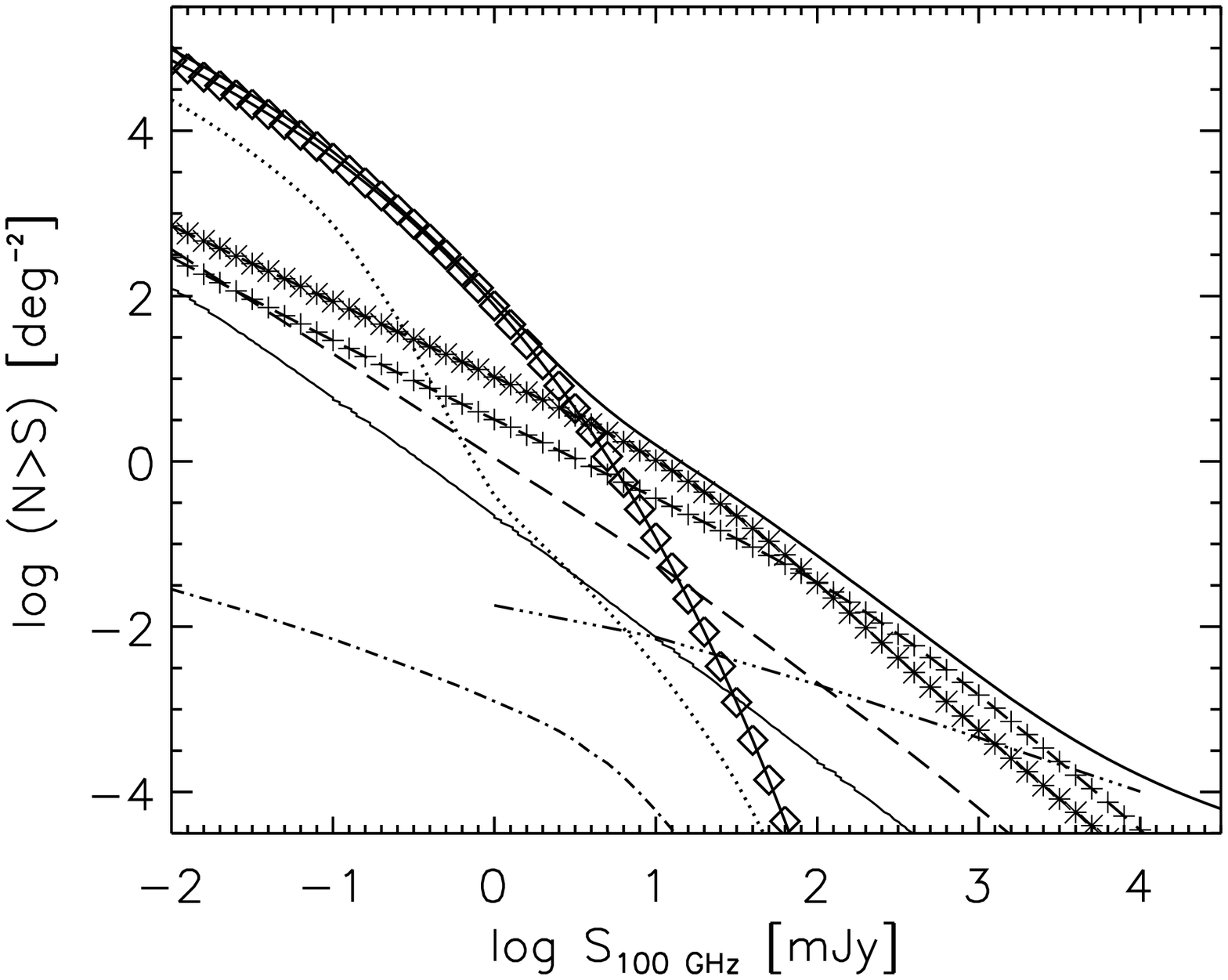} \\
\end{tabular}}
\caption{Predicted counts of galaxy-scale SZ effects (diamonds) at
30 and 100 GHz, compared with counts of flat-spectrum radio
quasars plus BL Lacs ($+$), flat-spectrum galaxies ($\ast$),
steep-spectrum radio sources (dashed line), advection-dominated
sources (thin solid line), normal plus starburst galaxies (dotted
line), GPS sources (De Zotti et al., 2000; three dots - dash), and
gamma-ray burst afterglows (Ciardi \& Loeb, 2000). }\label{counts}
\end{figure}

As discussed by De Zotti et al. (2003), these SZ signals may have
already shown up as the excess power on arcminute scales detected
by the BIMA experiment (Dawson et al., 2002). The forthcoming
sensitive high-frequency surveys might produce direct counts of
such signals. As shown by Fig.~\ref{counts}, we expect a surface
density of $\sim 0.3\,\hbox{deg}^{-2}$ at $S_{30{\rm GHz}}\simeq
1\,$mJy, and much higher counts at 100 GHz. Of course, the counts
at bright fluxes are likely dominated by SZ effects in clusters of
galaxies (not shown in Fig.~\ref{counts}), which however should be
easily distinguishable because of their much larger angular size
and much lower redshift.

\acknowledgements The roots of the work summarized in this paper
go back to 1977 when two of us (LD \& GDZ), under Alfonso's
guidance, started an investigation of the morphological,
dynamical, X-ray and microwave (SZ) properties of clusters of
galaxies in the framework of the then standard scenario for large
scale structure formation (adiabatic density perturbations in a
baryon-dominated universe), with special attention to the
observational windows (X-ray and microwave) whose exploration had
started only a few years before. One remarkable outcome of this
work was the demonstration of the possibility of determining
cluster distances directly, by combining X-ray and microwave
measurements. Remarkably, the work we are carrying out now, after
26 years, is much on the same line and, once again, we are
resorting to the SZ effect. Many thanks, Alfonso, for having
taught us such an effective research method!

\medskip\noindent
Work supported in part by MIUR and ASI.

\end{article}
\end{document}